\documentclass[twocolumn,showpacs,preprintnumbers,aps,prl]{revtex4}
\usepackage{graphicx}
\usepackage{dcolumn}
\usepackage{amsmath}
\usepackage{bm}

\begin{document}
\preprint{}

\title{Spectroscopic observation of the rotational Doppler effect}
\author{S. Barreiro$^1$, J.W.R. Tabosa$^2$, H. Failache$^1$, and
A. Lezama$^1$}

\affiliation{$^{1}$Instituto de F\'isica, Facultad de
Ingenier\'ia, C. Postal 30,
11000 Montevideo, Uruguay\\
$^{2}$Departamento de F\'isica, Universidade Federal de
Pernambuco, Cidade Universitaria 50670-901, Recife,PE, Brasil}

\date{\today}

\begin{abstract}
We report on the first spectroscopic observation of the rotational
Doppler shift associated with light beams carrying orbital angular
momentum. The effect is evidenced as the broadening of a Hanle/EIT
coherence resonance on Rb vapor when the two incident
Laguerre-Gaussian laser beams have opposite topological charges.
The observations closely agree with theoretical predictions.
\end{abstract}

\maketitle

\preprint{}

Light beams with twisted wavefronts, as is the case for the
Laguerre-Gaussian (LG) modes, are known to carry orbital angular
momentum (OAM) along their propagation direction. The properties
of such modes have attracted considerable attention in recent
years  and a wide range of applications were suggested
\cite{PADGETT04,ALLENBOOK03}. LG beams were used for atom trapping
and cooling \cite {KUGA97,WRIGHT00,PIRANDOLA03} and special
attention was put on the application of LG beams for the exchange
of orbital angular momentum between light and Bose-Einstein
condensates \cite {MARZLIN97,MARZLIN00,TEMPERE01,NANDI04}. Some of
us have demonstrated that OAM can be recorded in the position
dependent population and coherence of a cold atom sample
\cite{TABOSA99}, and transferred between internal atomic states.
In addition, the generation of new fields with OAM via non-linear
wave mixing in coherently prepared cold atoms was observed
\cite{BARREIRO03}. Quite recently, the use of photons in LG modes,
was suggested for quantum information processing. The state of
such a photon lies in a multi-dimensional Hilbert space,
describing the total (intrinsic plus orbital) angular momentum, in
which quantum computation with improved efficiency should be
possible \cite{BECHMANN00}. Entanglement between pairs of photons
in modes with OAM was recently reported \cite{VAZIRI03}.

The interaction of a moving atom with a LG field raises the
fundamental question of the Doppler effect \cite{ALLEN94}. As an
atom moves across the helicoidal wavefronts of the LG mode, it
experiences, in addition to the usual Doppler shift related to the
velocity in the light propagation direction (and a small shift
associated to radial motion in a curved wavefront), a most
intriguing frequency shift, the so called rotational Doppler
effect (RDE) associated to the azimuthal velocity. To date, the
RDE has only been observed interferometrically. The RDE results in
frequency shift when the light beam is rotated around its
propagation axis \cite{NIENHUIS96}. Such shift was observed in
\cite{COURTIAL98,COURTIAL98BIS} using millimeter-waves. In the
optical domain, a frequency shift in the field generated by a
rotating plate was observed in \cite{BASISTIY02}. A different
approach, used in \cite{BASISTIY03}, relates the RDE to the
asymmetric interferometric spatial pattern occurring in the
superposition of a Gaussian and a LG modes. In this work we
present the first experimental demonstration of the RDE arising
directly from the interaction of LG light beams with an atomic
sample.

Laguerre-Gaussian modes are usually identified with two integer
numbers: $l$ and $p$ \cite{ALLEN92}. The topological charge $l$
corresponds to the phase variation (in units of $2\pi $) of the
field along a loop encircling the optical axis. The integer $p+1$
corresponds to the number of  maxima of the field intensity along
a radius. In this study we are concerned with modes with $p=0$.
The Doppler shift experienced by a LG field in a moving frame is
given by \cite{ALLEN94}:
\begin{eqnarray}
\delta _{LG} =-\left[ k+\frac{kr^2}{2\left( z^2+z_R^2\right) }\left( \frac{%
2z^2}{z^2+z_R^2}-1\right) \right.   \nonumber \\
\left. -\frac{\left( 2p+\left| l\right| +1\right)
z_R}{z^2+z_R^2}\right] V_z-\left( \frac{krz}{z^2+z_R^2}\right) V_R
  -\left( \frac lr\right) V_\phi   \label{Doppler total}
\end{eqnarray}
where ($r,z,\phi $) and ($V_R,V_z,V_\phi $) represent in
cylindrical coordinates a position in space and corresponding
velocity of the moving frame. $z_R\equiv \pi w_0^2/\lambda $ is
the Rayleigh range and $w_0$  the beam waist.

The first term on the right hand side of Eq. \ref{Doppler total}
is the usual Doppler shift associated to motion along the
propagation axis direction, dominated by the leading term $-kV_z$
analogous to the Doppler shift of a plane wave with wavevector
$\mathbf{k}$ directed along $z$. The second term represents the
contribution to the Doppler of the radial velocity due to
wavefront curvature. This term is smaller with respect to the
first by a factor of the order of the beam divergence angle
$\theta $ (typically $\theta \sim 10^{-4}$ for well collimated
beams). The last term in Eq. \ref{Doppler total}, represents the
RDE which is proportional to the topological charge $l$ of the
mode. This term is smaller than the axial Doppler shift by a
factor $l\lambda /2\pi r$ which is of the order of $10^{-4}$ under
standard experimental conditions ($r\sim 1mm,\ \lambda \sim 1\mu
m$).

\begin{figure}
\includegraphics[width=2in]{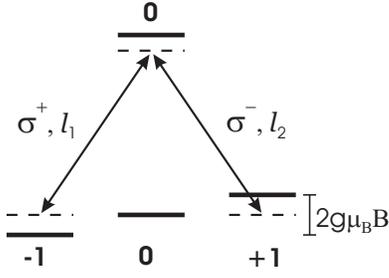}
\caption{\label{idea}Scheme used for the observation of rotational
Doppler effect on Hanle/EIT resonances illustrated on a
$F=1\rightarrow F^{\prime }=0$ transition. The two spatially
overlapped fields have identical frequency, opposite circular
polarizations and topological charges $l_1$ and $l_2$.}
\end{figure}

The observation of RDE through atom-field interaction was already
suggested in \cite{ALLEN94}. However the smallness of the RDE
compared to the axial shift makes the observation of the former
quite difficult under conditions where both contributions are
present. Our demonstration of RDE through atomic spectroscopy
relies on the observation of Doppler line broadening, due to RDE,
under conditions where other contributions (axial and radial) to
the Doppler broadening exactly cancel. The basic idea is presented
in Fig. \ref{idea}. We consider a transition between two atomic
levels, a ground level and an exited level with total angular
momentum $F$ and $F^{^{\prime }}=F-1$ respectively. Incident upon
this system, are two copropagating fields ($1$ and $2$) with
identical frequencies and opposite circular polarizations. The two
fields are LG modes with the same intensity, equal
beam waist $w_0$ and topological charges $l_1$ and $l_2$ respectively ($p=0$%
). For a moving atom, the resonance condition for a two-photon
(Raman) transition between ground state Zeeman sub-levels with
$\Delta m=\pm 2$, is given by: $\delta ^{^{\prime }}-\delta =0$
where $\delta ^{^{\prime }}=\delta _{LG}^1-\delta _{LG}^2$ and
$\delta $ represents the total Zeeman
shift ($\delta =2g\mu _BB$, $g$ is the ground state gyromagnetic factor and $%
\mu _B$ the Bohr magneton) induced by a static magnetic field $B$
oriented along the propagation axis. Since the axial and radial
contributions to the Doppler effect are the same for both fields
they exactly cancel and we are left with $\delta ^{^{\prime
}}=\left( \frac{l_1-l_2}r\right) V_\phi$. In
consequence, pure rotational Doppler line broadening should occur when $%
l_1\neq l_2$.

The two-photon resonance results in a reduction of the atomic
absorption around $B=0$ known as Hanle/EIT resonance
\cite{RENZONI97}. At low light levels, the homogeneous lineshape
$h$ of the Hanle/EIT resonance is given by \cite{SCULLYBOOK97}:
\begin{subequations}
\begin{eqnarray}
h\left( \delta ^{^{\prime }}-\delta ,r\right)  &=&AI_1\left(
r\right) I_2\left( r\right) L\left( \delta ^{^{\prime }}-\delta
\right)
\label{Linea homogenea} \\
L\left( x\right)  &=&\gamma /2\pi \left[ x^2+(\gamma /2)^2\right]
^{-1} \label{Lorentziana}
\end{eqnarray}
\end{subequations} where $A$ is a constant, $r$ is the atom radial
position and $I_j\left( r\right) $ [$j=1,2$] the intensity
distribution of field $j$. $\gamma $ is the relaxation rate of the
ground state atomic coherence. For an atomic vapor, the total
Hanle/EIT signal is given by:
\begin{equation}
S\left( \delta \right) =\int_0^\infty 2\pi rdr\int_{-\infty
}^{+\infty }h\left( \delta ^{^{\prime }}-\delta ,r\right) W\left(
V_\phi \right) dV_\phi   \label{forma de linea}
\end{equation}
where $W\left( V_\phi \right) $ is the velocity distribution at
thermal equilibrium at temperature T: $W\left( V_\phi \right)
=N(\pi /\alpha )^{-1/2}\exp \left( -\alpha V_\phi ^2\right) $ with
$\alpha =m/\left( 2k_BT\right) $ ($m$ is the atomic mass, $k_B$
the Boltzmann constant). From Eq. \ref{forma de linea} we see that
the lineshape corresponds to the homogeneous Lorentzian line when
$l_1=l_2$ and presents a distinctive
Doppler broadening for $l_1\neq l_2$. Expressing $V_\phi $ in terms of $%
\delta ^{^{\prime }}$ one gets:
\begin{eqnarray}
S(\delta ) &=&A\frac N{\sqrt{\pi /\alpha }}\int_{-\infty
}^{+\infty }d\delta ^{^{\prime }}\int_0^{+\infty
}I_1(r)I_2(r)L\left( \delta ^{^{\prime
}}-\delta \right)   \nonumber \\
&&\times \exp \left[ -\frac{\alpha \delta ^{^{\prime }2}}{(l_1-l_2)^2}%
\right] \frac{2\pi r^2}{l_1-l_2}d\,r
\end{eqnarray}
Finally, using the expressions for the far-field intensity of a LG mode, $%
I_j(r)=I_{0j}r^{2|l_j|}e^{-2r^2/w^2(z)}$ ($w\left( z\right) $ is
the propagation-distance dependent beam radius \cite{ALLEN94})\
one obtains:
\begin{equation}
S(\delta )=C\int_{-\infty }^{+\infty }L\left( \delta ^{^{\prime
}}-\delta \right) \,\left[ \frac{\alpha \delta ^{^{\prime
}2}}{(l_1-l_2)^2}+\frac 4{w^2(z)}\right] ^{-q}d\delta ^{^{\prime
}} \label{forma de linea final}
\end{equation}
where $q=|l_1|+|l_2|+3/2$ and $C$ is a coefficient depending on
$q$.

A simplified expression for $S(\delta )$ can be obtained in the
limit of a very narrow homogeneous resonance ($\gamma \rightarrow
0$):
\begin{equation}
S(\delta )=C\left[ \frac{\alpha \delta ^2}{(l_1-l_2)^2}+\frac
4{w^2(z)}\right] ^{-q}  \label{linea sin convolucion}
\end{equation}
and the full width at half maximum of the resonance is given by:
\begin{equation}
\Delta =\frac{4\left| l_1-l_2\right| }{\sqrt{\alpha }w\left(
z\right) }\sqrt{ 2^{(1/q)}-1}  \label{Delta }
\end{equation}
In the general case of a finite homogeneous line width, the
convolution integral \ref{forma de linea final} is evaluated
numerically. As indicated by Eqs. \ref{forma de linea final} and
\ref{Delta }, the width of the Doppler broadened resonance grows
as $\left| l_1-l_2\right| $ is increased.

\begin{figure}
\includegraphics[width=3.5in]{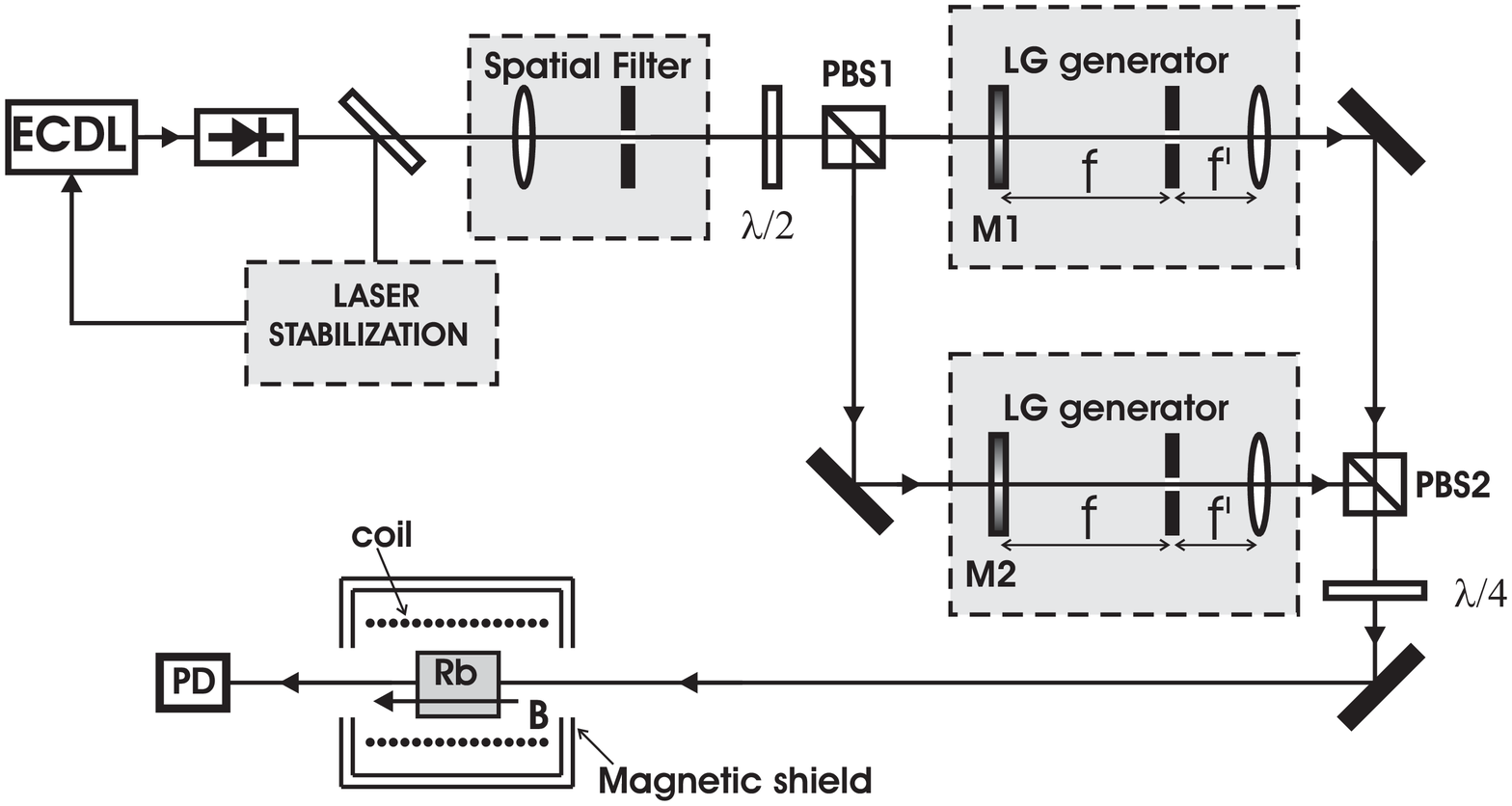}
\caption{\label{setup}Experimental setup. ECDL: extended cavity
diode laser. PBS: polarizing beam splitter, PD: photodiode. M1 and
M2 are the Laguerre-Gaussian mode-generating masks}
\end{figure}

We have observed the rotational Doppler broadening of the
Hanle/EIT signal on a $D1$ line transition of $^{87}$Rb in a
room-temperature vapor cell. The experimental setup is presented
in Fig. \ref{setup}. An extended cavity laser diode was used as
light source. The laser frequency was kept constant near the
center of the well resolved $5S_{1/2}\left( F=2\right) \rightarrow
5P_{1/2}\left( F=1\right) $ atomic transition. A spatial filter
was used to improve the beam profile. Using a polarizing beam
splitter, the light beam was divided into two equal-intensity arms
with orthogonal linear polarizations. Each beam was sent to a LG
mode generator \cite{BASISTIY02} consisting on a mask imprinted on
a photographic film, a pinhole placed at the first order focus of
the mask and a collimating lens with focus on the pinhole. The
masks were computer-generated binary spiral zone plates \cite
{HECKENBERG92} recorded on high contrast photographic film. The
two LG mode generators were identical except for the relative
orientation of the mask
allowing the generation of fields with equal or opposite topological charge $%
l$. After propagating equal distances, the two fields were
recombined in a second polarizing beam-splitter. A quarter-wave
plate placed after the beam-splitter transforms the polarizations
of the two fields into circular and opposite. The overlapping
fields are sent through the atomic gas cell placed inside a
two-layer $\mu $-metal shield. Inside the shield, a solenoid
controls the longitudinal magnetic field. The total light
intensity transmitted through the sample is monitored with a
photodiode as a function of the applied magnetic field. The light
power at the atomic sample was controlled with neutral density
filters. During measurements the light power at the sample was
approximately  1 $\mu W$ (beam radius: $0.5\ -\ 0.9\ mm$, see
below).

To improve sensitivity, a small AC modulation was added to the DC
value of the magnetic field and lock-in amplification was used for
the photodiode current. The signal recorded as a function of $B$
corresponds to the derivative of the Hanle/EIT signal (provided
that the AC modulation is small and slow enough). We measure the
linewidth as the peak-to-peak distance on the derivative signal.

To demonstrate the occurrence of Doppler broadening due to RDE we
have systematically compared the resonance width obtained under
identical experimental conditions when the two fields acting on
the atomic sample posses equal or opposite topological charges. In
the  case $l_1=l_2$ no RDE is expected and the observed resonance
corresponds to the homogeneous lineshape. When $l_1=-l_2$,  RDE\
line broadening should be present. The two situations can be
obtained in our experimental setup by using identical masks on the
two arms of the light pass. If the two mask are equally oriented
we get $l_1=l_2$. Reversing one of the masks results in
$l_1=-l_2$.

An important requirement for our demonstration is to ensure that
no spurious
line broadening is introduced due to misalignment. Indeed, a small angle $%
\varepsilon $ between the axis of the two LG modes results in
Doppler broadening of the order of $\varepsilon \Delta _{Dopp}$
($\Delta _{Dopp}\simeq 500$ MHz). By slightly misaligning the
overlapping of the two fields, we have measured a linewidth
increase of 110 KHz per milliradian of angular misalignment. Our
beam overlapping procedure was accurate to better than 0.2 mrad.
In consequence the angular misalignment contribution to the
observed width was less than 22 KHz. A confirmation of this is
reached by comparison of the linewidth obtained with $l_1=l_2$ and
with a single LG field linearly polarized. In the latter case, the
two fields participating in the Hanle/EIT resonance are the
circular components of the single field and no broadening due to
misalignment is possible. The two situations gave the same
linewidth ($\sim 52$ KHz at low light power). Finally, we have
checked that no spurious broadening was introduced by the
reversing of the orientation of one mask to change the topological
charge sign. For this, a pair of masks with $l_1=l_2=0$ was used.
No change in the linewidth was observed after reversing the
orientation of either of the two masks.

\begin{figure}
\includegraphics[width=3.5in]{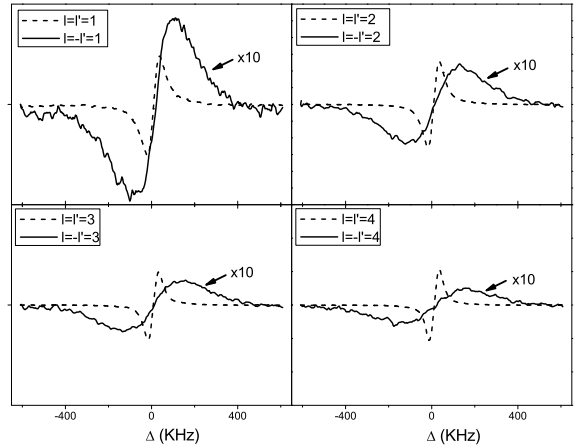}
\caption{\label{curvas experimentales}Observed Hanle/EIT
resonances as a function of the total Zeeman shift $%
\delta =2g\mu _BB$ for different pairs of LG mode-generating
masks. Dashed(solid): same(opposite) topological charge on each
field.}
\end{figure}

\begin{figure}
\includegraphics[width=3.5in]{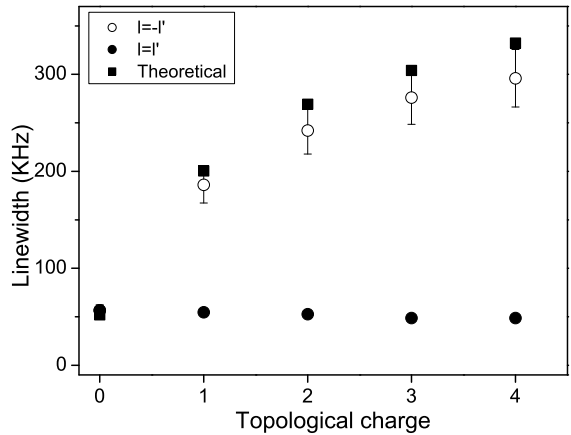}
\caption{\label{anchos}Linewidht dependence on topological charge.
Circles experimental results: solid(hollow) equal(opposite) charge
on each field. Triangles: theoretical prediction.}
\end{figure}

In Fig. \ref{curvas experimentales} we present the recorded signal
for different pairs of LG-mode-producing masks with $\left|
l_1\right| =\left| l_2\right| =l=1,2,3,4$. All four cases produce
narrow resonances of similar
width for $l_1=l_2$. A visible broadening due to RDE\ is observed for $%
l_1=-l_2$. A variation of the signal level with $l$ is also
observed. Such variation is expected as a result of the variation
of the radial light intensity-distribution for LG modes with
different $l$. The dependence of the peal-to-peak linewidth with
$l$ is shown in Fig. \ref{anchos}. Almost constant width is
observed for $l_1=l_2$ with a slight increase for small $l$ due to
intensity broadening. A monotonic growth of the linewidth with $l$
is observed when $l_1=-l_2$. Also shown in Fig. \ref{anchos} are
the calculated width obtained using Eq. \ref{forma de linea final}
with a homogeneous width of $52$ KHz. In the calculations, we used
$w\left( z\right)=0.5,\ 0.65,\ 0.74,\ 0.83,\ 0.89\ mm$ for
$l=0,1,2,3,4$ respectively. These values were taken from the fit
of the experimental beam cross section (recorded with a CCD
camera) to the theoretical intensity profile of a LG mode. They
increase with $l$ as expected \cite{KOTLYAR06}.

The results presented in Figs. \ref{curvas experimentales} and
\ref{anchos} are a clear indication of rotational Doppler
broadening on the Hanle/EIT resonances produced by two LG fields
with different topological charges. A good agreement is observed
between the theoretically predicted and the experimentally
observed variation of the resonance width with $l$. We notice that
the experimentally observed width are slightly smaller than the
predicted ones. We attribute this fact to small contamination of
the actual light field with a Gaussian mode \cite{VAZIRI02}
($l=0$).

In summary, we have experimentally observed the rotational Doppler
shift associated with light beams carrying OAM via the coherent
interaction between these beams and an ensemble of rubidium atoms
at room temperature. The effect is clearly evidenced as an
inhomogeneous broadening of the Hanle resonance depending on
atomic azimuthal velocity. These observations constitute the first
demonstration of light-atom interaction depending on azimuthal
atomic velocity. They provide experimental support to the
suggested use of two-photon coherent resonances for the
manipulation of the rotational motion of atomic samples
\cite{MARZLIN97,MARZLIN00,NANDI04}.

This work was supported by Fondo Clemente Estable, CSIC and
PEDECIBA (Uruguay). J.W.R.T. acknowledges financial support from
CNPq (Brazil).

\end{document}